# T$_2$ Relaxation during Radiofrequency (RF) pulses

**Peder E. Z. Larson**

Department of Radiology and Biomedical Imaging, University of California, San Francisco, CA, 94143, USA

Corresponding Author: Peder Larson (Peder.Larson@ucsf.edu)

**Index**

## 1 INTRODUCTION

Radiofrequency (RF) pulses are a critical part of every MRI pulse sequence, and must be specifically designed for ultrashort echo time (UTE) and zero echo time (ZTE) acquisitions. When considering the behavior of RF pulses, most often longitudinal T$_1$ or transverse T$_2$ relaxation is assumed to be negligible during the RF pulses themselves. This is usually valid with conventional sequences since most tissue T$_1$s and T$_2$s are much longer than typical RF pulse durations. However, when imaging tissues that have transverse relaxation times that are of the order of, or shorter than, the RF pulse duration, as is often the case with UTE and ZTE MRI, then relaxation during the pulse must be considered. This chapter covers the theory of $T_2/T_2^*$ relaxation during an RF pulse, and the implications as well as applications of this for imaging of short- and ultrashort-$T_2^*$ species.

## 2   THEORY

To determine the effect of relaxation during an RF pulse, we simply need to use the Bloch equations. However, solutions of the Bloch equations used in RF pulse simulation and design, such as the Shinnar-Le Roux (SLR) transform, cannot be used since they neglect relaxation. Numerical solutions to the Bloch equation can be used to provide the most accurate simulations of RF pulse profiles.

However, to gain insight into the interaction between an RF pulse including longitudinal and transverse relaxations, the following approximate solution to the Bloch equation for the longitudinal magnetization, $M_Z$, is useful:

$$M_Z(T_2) \approx M_0(1 - T_2 \int_{-\infty}^{\infty} |\Omega_1(f)|^2 df) \tag{1}$$

Here, $M_0$ is the equilibrium magnetization, $T_2$ is the transverse relaxation time, $f$ is frequency, and $\Omega_1(f)$ is the Fourier transform, or frequency spectrum, of the RF pulse. This result was derived assuming $T_2$ is short relative to fluctuations in the RF pulse shape and using a small-tip approximation [1]. This shows that there is a trade-off between total RF spectral power (e.g., pulse bandwidth) and short-$T_2$ signal attenuation.

The effect of RF pulses on short-$T_2$ components can also be understood in terms of spectral linewidths (Fig. 1). $T_2$ is inversely proportional to the linewidth, meaning that short-$T_2$ species have broad linewidths and long-$T_2$ species have narrow linewidths. The overlap between the spectrum of the RF pulse and the tissue spectrum determines the approximate degree of excitation. The narrow spectrum of long-$T_2$ species is more easily covered by the RF spectrum and thus they are easily excited [2]. Broad short-$T_2$ species require a wide bandwidth RF pulse to be fully excited. Thus, a narrow bandwidth RF pulse can fully excite longer $T_2$ species but only partially excites shorter $T_2$ species (Fig. 1).

This intuitive understanding can be extended to off-resonance situations. For example, magnetization transfer (MT) applies RF pulses far from the water resonance frequency, which only leads to excitation of very broad linewidth ultrashort-$T_2$ components. Fat suppression pulses are also applied at the lipid resonance frequencies with relatively narrow bandwidths in order to selectively excite fat but not long-$T_2$ water.

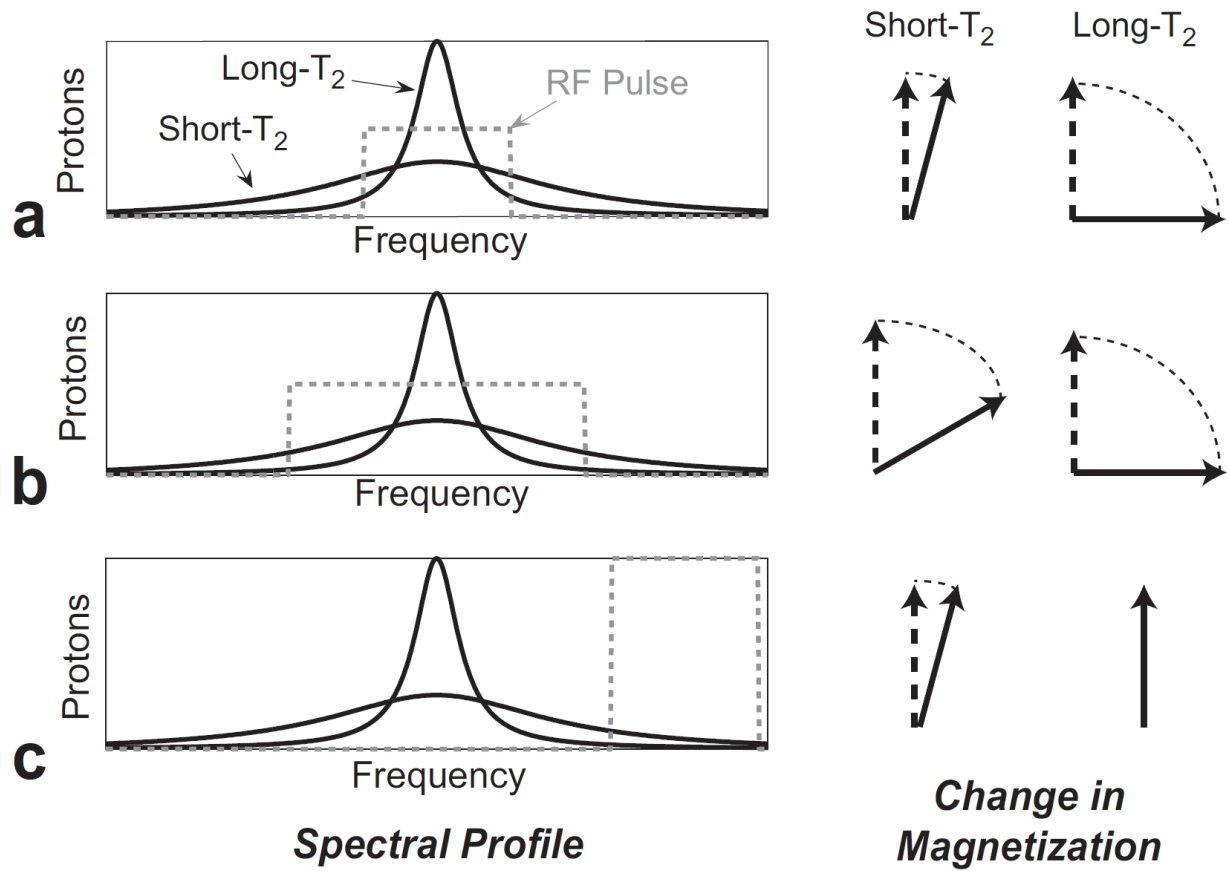

Figure 1. Illustration of how spectral linewidth (proportional to $T_2$) and RF pulse bandwidth determine excitation. (a) Narrow bandwidth (dashed line) RF pulse that overlaps the majority of the long-$T_2$ spectrum but only a small fraction of the short-$T_2$ spectrum. The long-$T_2$ species is excited more by the pulse than the short-$T_2$ species. (b) Wide bandwidth (dashed line) RF pulse that overlaps both the long- and short-$T_2$ spectra, thus exciting both. (c) This also applies to MT pulses, which are applied off-resonance primarily to excite ultrashort-$T_2$ species. (Reproduced with permission from Ref. [1])

## 3    EXCITATION AND $T_2$ RELAXATION

The purpose of excitation is to generate transverse magnetization for imaging. When imaging ultrashort-$T_2$ species, the major consideration is to ensure there is an adequate flip angle applied to the rapidly decaying component. As shown in Fig. 2, the flip angle decreases with shorter $T_2$s, causing noticeable decreases when $T_2$ is of the order of, or shorter than, the pulse duration. Also note that shorter $T_2$s also lead to blurring of the slice profile.

To achieve sufficient excitation of an ultrashort-$T_2$ component, the intuition from Fig. 1 tells us that RF pulses should be designed to have as large a bandwidth as possible, as shown in Fig. 3. Equivalently, for a given pulse shape the RF pulse should be as short as possible, since bandwidth scales inversely with duration. Ultimately, shortening the RF pulse is limited by the MR system peak $B_1$ amplitude and peak gradient strength (when performing slice selection).

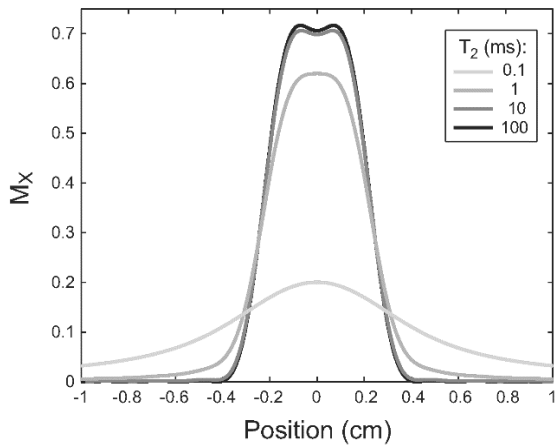

Figure 2. Bloch equation simulated slice profile versus $T_2$ for a 1 ms, 45° half-pulse [3, 4]. $T_2$ relaxation during the pulse blurs the desired 5 mm slice profile and decreases the flip angle for the shortest $T_2$ values.

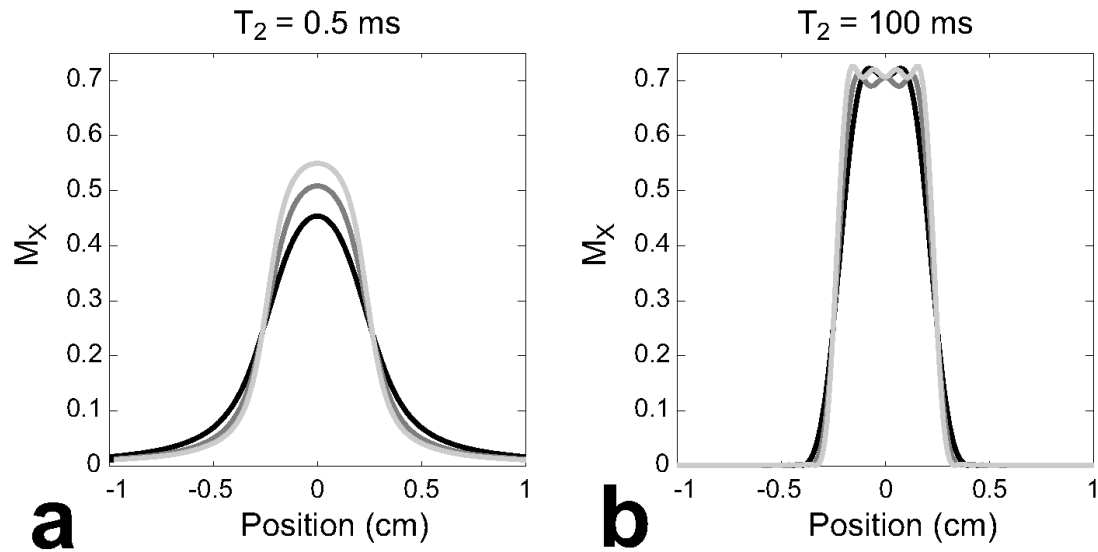

Figure 3. Simulated half-pulse slice profiles for 45°, 2 ms pulses with bandwidths of 1.2 kHz (black), 1.8 kHz (dark gray), and 2.4 kHz (light gray line). As expected, the increasing bandwidth increases the short-$T_2$ excitation flip angle (a), but only changes the profile sharpness of long-$T_2$ components (b).

Changes in excitation as a function of $T_2$ have also been exploited to improve the contrast for short-$T_2$ components [5]. This approach uses an acquisition with a short RF pulse, which has high signal from both long- and short-$T_2$ components, and an acquisition with a long RF pulse, which has high signal only from long-$T_2$ components. The long-$T_2$ component signal can then be suppressed by subtracting these two images, leading to improved short-$T_2$ contrast (Fig. 4).

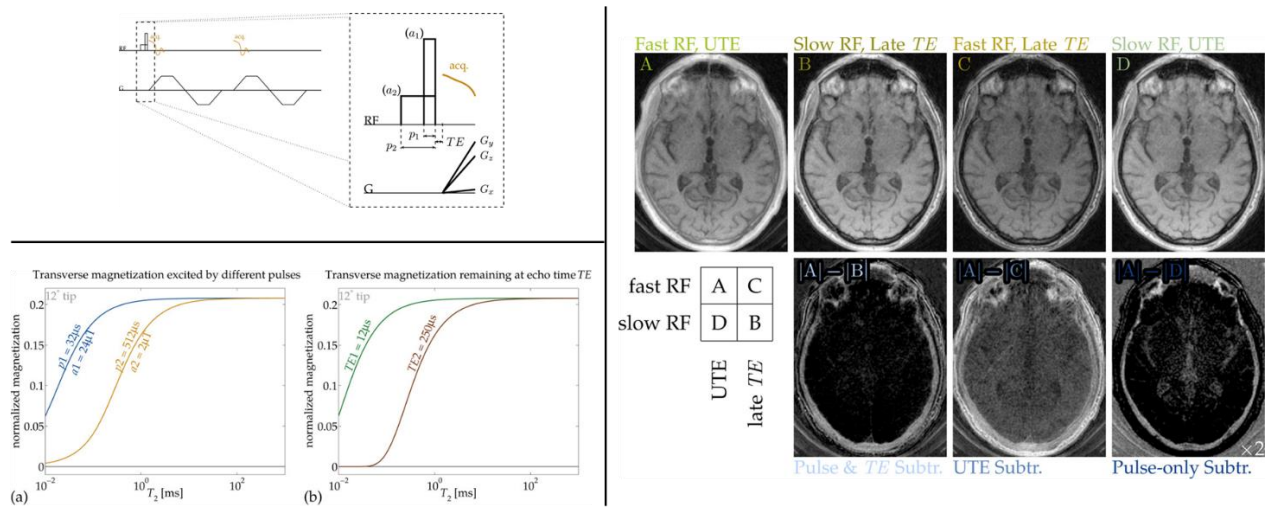

Figure 4. RF excitation pulses of different lengths (top left) can be used to create $T_2$ selectivity (bottom left). For comparison, the $T_2$ selectivity at different TEs is also shown. The images (right) show how the fast and slow RF pulse images as well as different TE images can be subtracted to create short-$T_2$ contrast, in this case providing excellent depiction of cortical bone in the skull which has a $T_2^* \sim 0.3$ ms. (Adapted with permission from Ref. [5])

Short-$T_2$ component excitation has also been combined with fat suppression in a soft-hard composite pulse approach [6]. In this approach, the composite pulse contains a narrow bandwidth soft pulse centered on the fat peak with a small negative flip angle ($-\alpha$) and a short rectangular pulse with a small positive flip angle ($\alpha$). The fat magnetization experiences both tipping-down and -back with an identical flip angle and thus returns to its equilibrium state, leaving only the water magnetization excited. This avoids short-$T_2$ component saturation that happens during conventional fat saturation RF pulses (Fig. 5).

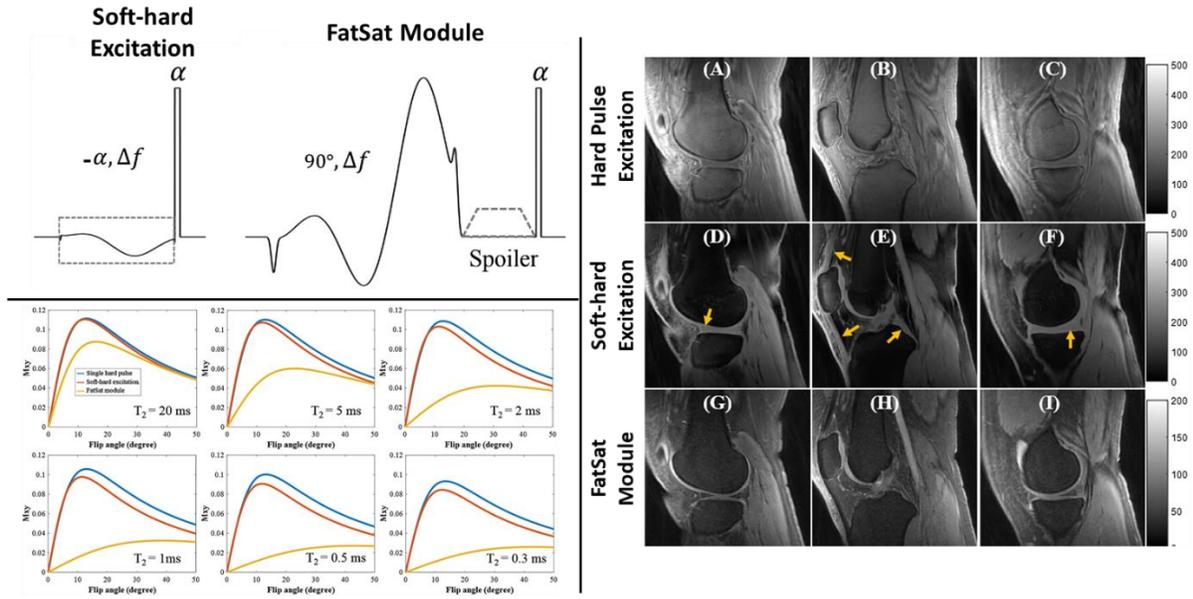

Figure 5. Short-$T_2$ selective excitation can be achieved with a composite soft-hard excitation. (Top left) This consists of a narrow bandwidth negative flip angle ($-\alpha$) pulse followed by high bandwidth pulse with a positive flip angle ($+\alpha$) of the same magnitude. The simulations (bottom left) and in vivo knee imaging results (right) show that this strategy creates improved short-$T_2$ component contrast without fat signal compared to standard 3D UTE with hard pulse excitation, and also has less suppression of water short-$T_2$ components compared to use of a FatSat module as highlighted by the orange arrows pointing to meniscus, ligaments, and tendons. (Adapted with permission from Ref. [6])

## 4 SATURATION AND $T_2$ RELAXATION

The $T_2$ selectivity of RF pulses has also been exploited to design saturation pulses intended to selectively saturate long-$T_2$ components, and provide improved contrast for short-$T_2$ components. One challenge in imaging short-$T_2$ components is that they often have smaller signals due to relaxation and/or lower proton density compared to long-$T_2$ components. This problem can be overcome by designing low bandwidth 90° flip angle RF pulses, which are followed by a spoiling gradient to suppress spins excited by the pulse. Using the intuition in Fig. 1a, these low-bandwidth pulses overlap the entire long-$T_2$ linewidth, causing full 90° excitation, while short- and ultrashort-

$T_2$ components have smaller or incomplete excitation, and thus will not be suppressed by the resulting spoiler gradient (Fig. 6).

The disadvantages of long-$T_2$ suppression pulses are that they are inherently sensitive to off-resonance, as they rely on use of a narrow pulse bandwidth for contrast, and, like other suppression pulse schemes, are sensitive to RF field inhomogeneities that lead to imperfect flip angles.

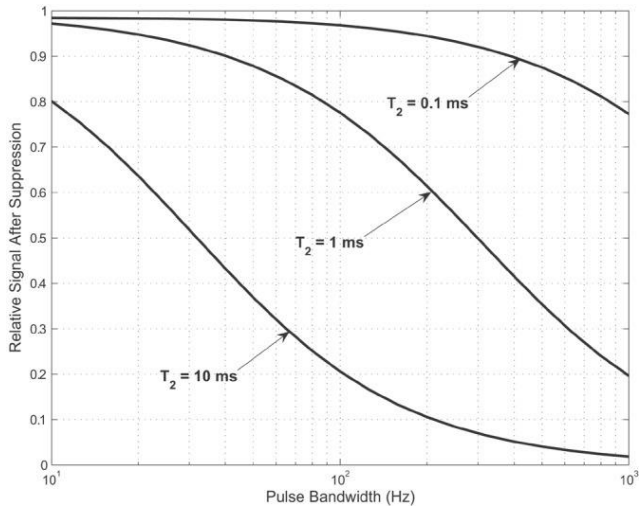

Figure 6. Simulation results for longitudinal magnetization remaining after suppression pulses of different bandwidths for various $T_2$ values. As the pulse bandwidth increases, the signal remaining after suppression decreases. As $T_2$ increases, the signal also decreases. This simulation is consistent with the intuition shown in Fig. 1. (Reproduced with permission from Ref. [1])

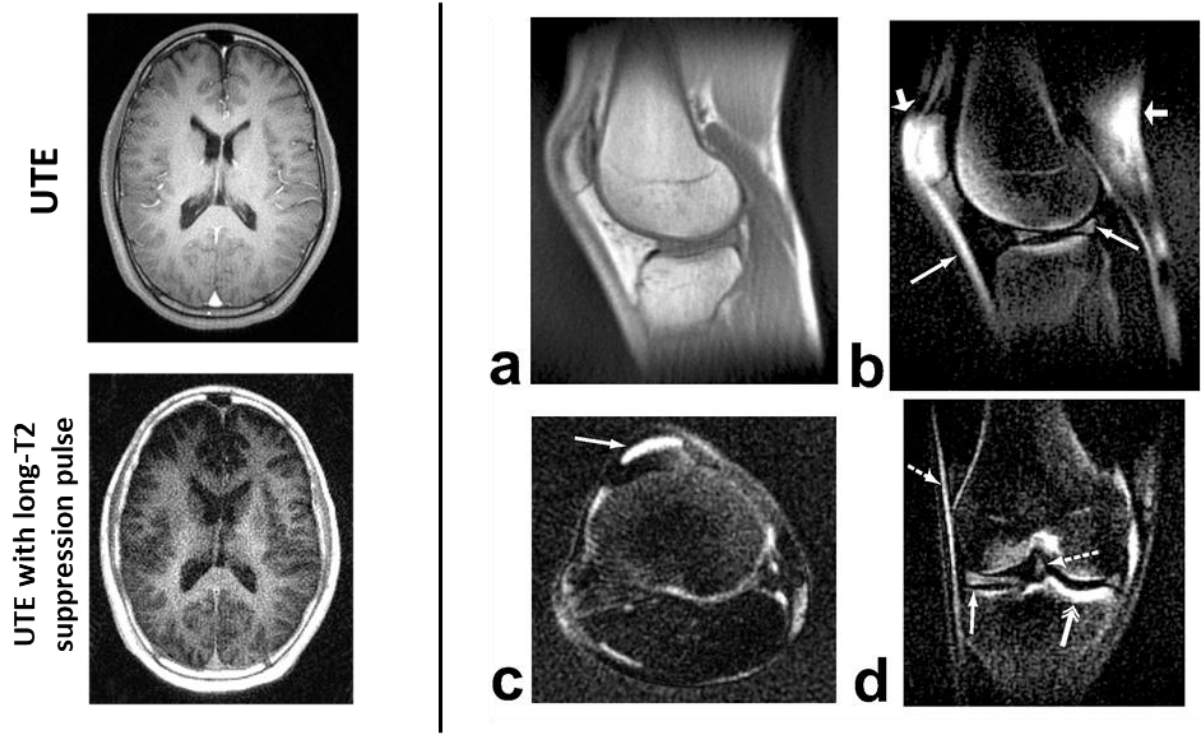

Figure 7. Long-$T_2$ signal suppression pulse results in vivo. (left) In the brain, a low bandwidth saturation pulse applied at the water resonance frequency creates improved contrast for short-$T_2$ components in myelin as well as connective tissue and cortical bone compared to UTE without long-$T_2$ signal suppression. (right) 3D UTE results in the knee show greatly improved contrast of short-$T_2$ structures when using a dual-band suppression pulse (b-d) to suppress long-$T_2$ water and fat components compared to no suppression in (a). This provides excellent visualization of the menisci, patellar tendon, iliotibial band, and anterior cruciate ligaments, as well as the thickened tibial cortex. There are some fat suppression failures (short arrows in b) due to $B_0$ field inhomogeneity. (Adapted with permission from Ref. [1])

The principles of long-$T_2$ suppression pulses can also be applied with fat suppression, which is especially important for musculoskeletal applications where short-$T_2$ tissues such as tendons, ligaments, cartilage, and bone are often adjacent to fat. One approach that can be used is to create "dual-band" RF pulses, a single RF pulse which includes relatively narrow saturation bandwidths at both fat and water resonances to selectively excite long-$T_2$ components at these two resonances [1].

Another approach is to use separate 90° saturation pulses at both the fat and water resonances, as illustrated in Fig. 7. This approach also includes low bandwidth refocusing 180° pulses on alternate scans. These are included to correct for imperfections in the flip angle of the 90° saturation pulses, which is a challenge for 90° RF pulse based saturation schemes. By using low bandwidth pulses, only the long-$T_2$ components experience this refocusing (Fig. 8).

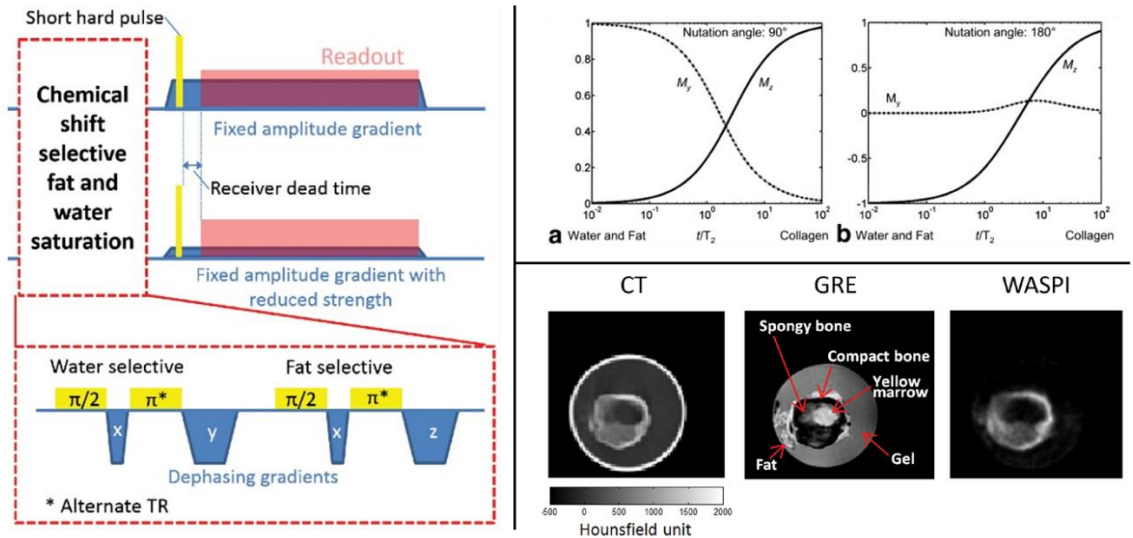

Figure 8. (left) The water- and fat-suppressed proton projection imaging (WASPI) sequence. This uses separate water and fat selective low bandwidth suppression pulses prior to a ZTE readout to selectively suppress long-$T_2$ components at both fat and water resonances. The 180° pulses provide robustness to flip angle errors. (top right) Simulated magnetization vs. $t/T_2$ following perfect rectangular RF pulses. (a) For low bandwidth 90° pulses, only the long-$T_2$ components are rotated into the transverse plane (b) Similarly, for the low bandwidth 180° pulses, only the long-$T_2$ components are inverted. (bottom right) Using this tissue suppression scheme, the long-$T_2$ water gel and long-$T_2$ fat components are suppressed, leaving a selective image of cortical bone with contrast similar to CT. (Adapted with permission from Refs. [7, 8])

$T_2$ selective suppression can also be achieved in a manner similar to MT by applying off-resonance contrast ("UTE-OSC") RF pulses [9]. Following the off-resonance RF pulse, only the short-$T_2$ components are excited (Fig. 1c), and these are suppressed by a spoiler gradient. The off-resonance saturated image is then subtracted from an unsuppressed image, leading to improved contrast of short-$T_2$ components. This approach is the inverse of long-$T_2$ suppression pulses (Fig. 9).

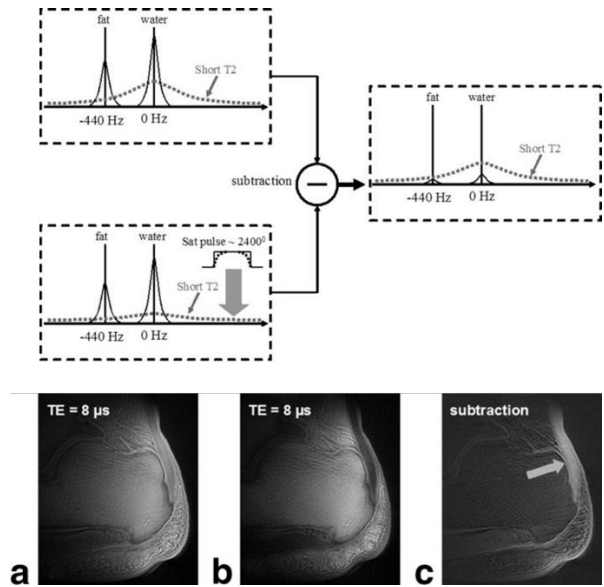

Figure 9. UTE with off-resonance saturation contrast (UTE-OSC). This includes two acquisitions with and without an off-resonance saturation pulse. The saturation pulse was placed +1 to +2 kHz away from the on-resonance water peak to minimize its spectral overlap with that of long-$T_2$ water and fat. Short-$T_2$ species have a broad spectrum and are suppressed by the saturation pulse. Subtraction of UTE images with and without off-resonance saturation pulse suppresses long-$T_2$ water and fat signals, leaving high contrast for short-$T_2$ species. (bottom) This is clearly shown in the ankle UTE imaging result, which shows unsaturated (a) and off-resonance saturated (b) images, from which subtraction leads to high contrast visualization of the Achilles' tendon (arrow in c). (Adapted with permission from Ref. [9])

## 5 INVERSION AND $T_2$ RELAXATION

Inversion pulses are valuable tools that are typically used to create $T_1$ contrast by including a time delay after a 180° pulse, also known as inversion recovery (IR). Using the principles in this chapter, we can also integrate $T_2$ selectivity into inversion pulses for short-$T_2$ imaging applications.

One approach to do this is to use low bandwidth inversion pulses to prepare the magnetization and then combine with non-inverted images [10]. Since the low bandwidth pulses primarily invert long-$T_2$ components, they are suppressed in the combined images and provide improved short-$T_2$ contrast. This strategy can also be extended to include fat suppression by combining long-$T_2$ fat-inverted and long-$T_2$ water-inverted images (Fig. 10).

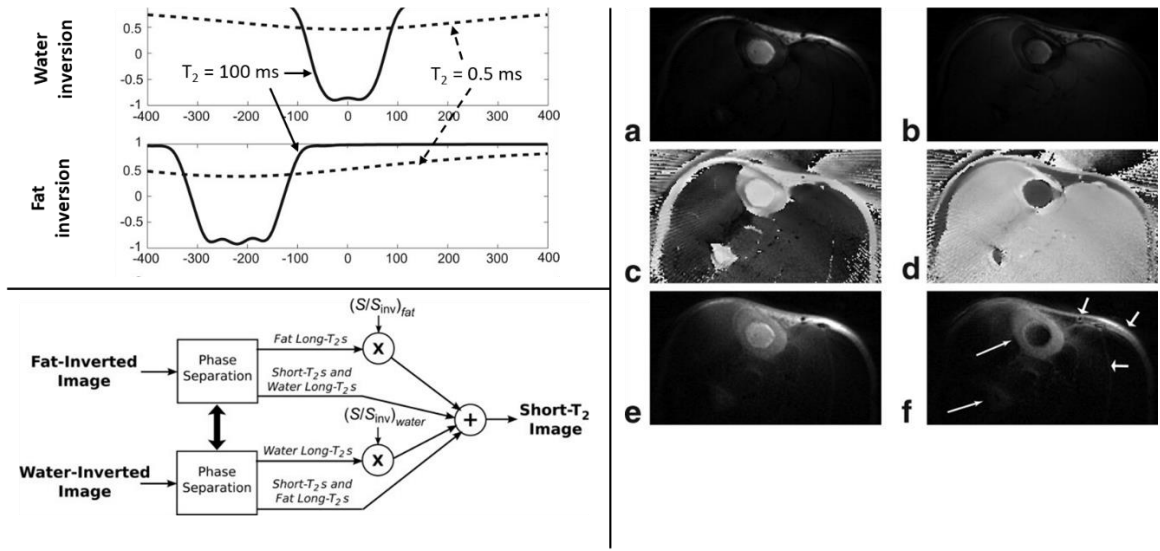

Figure 10. (top left) Adiabatic inversion pulses can create $T_2$ selective contrast. Simulations were performed for 30 ms Lorentz pulses. When the two images are combined, the long-$T_2$ species at both resonances are suppressed. The short-$T_2$ species, while attenuated, have significantly improved contrast. For the values shown, the short-$T_2$ to long-$T_2$ contrast is increased by a factor of 10. (bottom left) For fat and water long-$T_2$ suppression, fat-inverted and water-inverted images are acquired, and then separated into inverted and non-inverted components based on the corrected phase. (right) Axial images at the middle of the lower leg: (a) water-inverted magnitude image; (b) fat-inverted magnitude image; (c) water-inverted phase image, corrected; (d) fat-inverted phase image, corrected; (e) sum of (a,c) and (b,d); (f) scaled sum of (a) and (b), showing improved contrast of cortical bone as well as skin and signal between muscles and fascicles. (Adapted with permission from Ref. [10])

Short-$T_2$ imaging can also be combined with IR to suppress long-$T_2$ components [11]. This can simply be based on $T_1$ selectivity, but by designing reduced bandwidth inversion pulses this can improve the contrast for short-$T_2$ components. This has been successfully applied with both single and dual IR (DIR) UTE sequences [12–15]. As illustrated in Fig 11, the low bandwidth inversion pulses in these approaches aim to invert long-$T_2$ component longitudinal magnetization while minimally perturbing short-$T_2$ component magnetization. TIs are then chosen to null long-$T_2$ components based on their $T_1$s (Fig. 11).

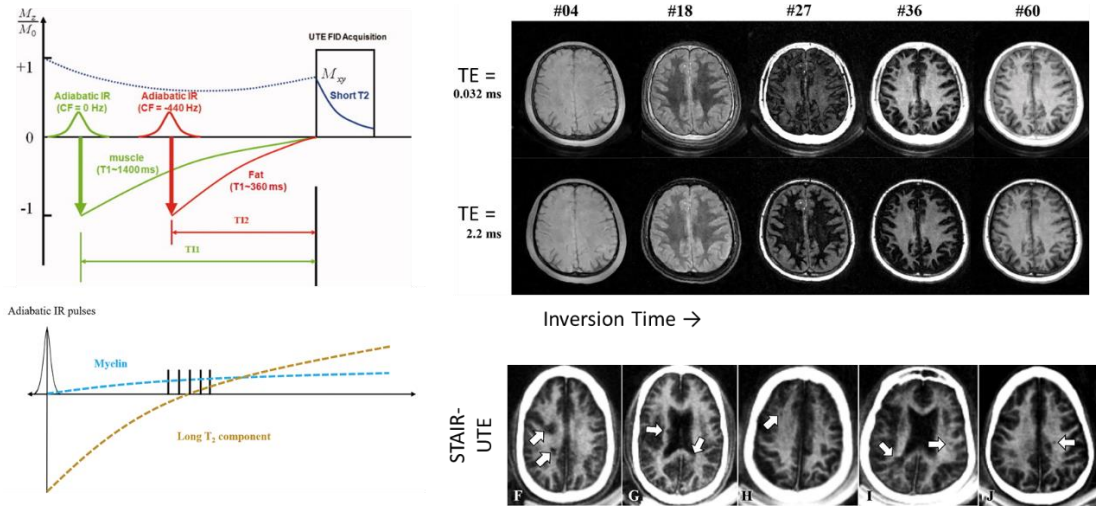

Figure 11. (left) Two examples of the expected signal response in IR-UTE. The short-$T_2$ components (e.g., myelin protons) are partially tipped by the inversion pulse whereas the long-$T_2$ components are fully inverted. At the chosen TI, the long-$T_2$ components are nulled while the short-$T_2$ components have experienced $T_1$ recovery. (right) Example results of IR UTE in the brain show how the contrast varies with TI and TE. Combining echo subtraction with a well chosen TI leads to short-$T_2$ selective images of myelin, shown using the STAIR-UTE method in a multiple sclerosis patient with demyelinating lesions (arrows). (Adapted with permission from Refs. [12–14])

# 6 CONCLUSION

For short-$T_2$ imaging with UTE and ZTE MRI pulse sequences it is important to understand the effects of $T_2$ relaxation during RF pulses. This becomes significant when $T_2$ is of the order or shorter than the RF pulse duration. The main intuition presented in this chapter is that the $T_2$ selectivity is determined by the overlap of the signal component linewidths with the RF pulse frequency profile. The linewidth is inversely proportional to $T_2$, while the RF profile is determined by the pulse bandwidth and frequency offset. This chapter shows how to ensure adequate excitation of short-$T_2$ components as well as methods for improving short-$T_2$ component contrast through the use of T$_2$ selective excitation, saturation and inversion RF pulses. The chapter also describes how fat suppression can be integrated and optimized for short- and ultrashort-T$_2$ imaging.